\documentclass{article}
\usepackage{dcase2017,amsmath,graphicx,url,times,booktabs, tabularx}

\title{Ensemble of deep neural networks for acoustic scene classification}

\name{Venkatesh Duppada, Sushant Hiray}
\address{Seernet Technologies, LLC \\
\{venkatesh.duppada, sushant.hiray\}@seernet.io}
\begin{document}

\ninept
\maketitle

\begin{sloppy}
	
	\begin{abstract}
		Deep neural networks (DNNs) have recently achieved great success in a multitude of classification tasks. Ensembles of DNNs have been shown to improve the performance. In this paper, we explore the recent state-of-the-art DNNs used for image classification. We modified these DNNs and applied them to the task of acoustic scene classification. We conducted a number of experiments on the TUT Acoustic Scenes 2017 dataset to empirically compare these methods. Finally, we show that the best model improves the baseline score for DCASE-2017 Task 1 by 3.1\% in the test set and by 10\% in the development set.
	\end{abstract}
	
	\begin{keywords}
		Deep learning, Convolutional Neural Networks, Recurrent Neural Networks, Ensemble 
	\end{keywords}

	\section{Introduction}
	\label{sec:intro}
	
	Acoustic scene classification(ASC) \cite{barchiesi2015acoustic} focuses on recognizing various environmental sounds from a sound sample. The broader category of Computational Auditory Scene Analysis aims to model the human auditory system and its mechanisms to detect, identify, separate and segregate sounds in the same way that humans do\cite{wang2006computational}. Amongst the existing acoustic classification tasks, ASC is a challenging task since various environmental sounds have similar background noises and span a wider range of frequencies. ASC is particularly interesting to researchers due to its wide applications in audio tagging \cite{cai2006flexible}, audio indexing using wearable devices \cite{shah2012lifelogging},  robot navigation systems \cite{chu2006scene} etc.
	
	Image classification and detection have been remarkably popular
	in recent years. The popularity is attributed primarily to the availability of large annotated standard datasets \cite{deng2009imagenet}. Audio classification and detection have not attracted a similar level of attention. The DCASE challenge is a step towards creating a standard dataset for researchers. The 2013 iteration \cite{giannoulis2013detection} included challenges for scene classification and synthetic acoustic classification. The DCASE Challenge 2016 \cite{heittola2016dcase2016} comprised of 4 tasks: ASC, Acoustic Event Detection, Sound Event Detection in Real Life Audio and Domestic Audio Tagging. The DCASE Challenge 2017 \cite{DCASE2017challenge} comprises of 4 tasks: ASC, detection of rare sound events, sound event detection in real life audio, large scale weakly supervised sound event detection for smart cars.
	
	In this work, we focus on the challenge 1 of DCASE Challenge 2017. The goal of ASC is to classify a test recording into one of the provided 15 predefined classes that characterizes the environment in which it was recorded. This challenge is in continuation of the previous year's challenge with additional data. 
	
	Early work in this area focused primarily on using classifiers such as GMM-HMM \cite{chum2013ieee}, tree bagger classifiers \cite{olivetti2013wonders}, support vector machines \cite{geiger2013large}. Some of these classifiers cannot effectively model temporal dynamics of audio. With the recent advancements in deep learning, many new DNN architectures have been studied which are better at encoding the temporal nature. Also DNNs are better at abstracting the large feature sets, usually associated with audio clips. Quite a few teams experimented with various deep learning approaches in the previous challenge. The challenge winners\cite{eghbal2016cp} for DCASE 2016 extracted MFCC and i-vectors and used Deep Convolutional Neural Networks (DCNN) to improve the performance over conventional baseline model. \cite{valenti2016dcase} proposed a vanilla DCNN with a fully connected layer at the end, whereas \cite{lidy2016cqt} pre-processed the audio clips using CQT before using a DCNN. \cite{bae2016acoustic} experimented with a parallel combination of CNN and LSTM for the final classification. 
	
	In this work, we apply various state-of-the-art DCNNs from image classification tasks. We modified these architectures to apply them on the audio dataset for ASC. Specifically, we use the following architectures: (1). LeNet (2). SqueezeNet (3). 1-D CNN. We also tried some deeper networks such as: Highway Networks, Densely Connected CNN. However, due to the lack of availability of a large annotated dataset, these variants didn't perform well. We also experimented with multiple types of feature extractors: extracting log-mel spectrograms by varying the hop-lengths and the sampling frequency. We compare the results from these DCNN models with the baseline DNN model. Finally, we also show that creating an ensemble of these networks gives a much better performance on the fore mentioned task, effectively improving the baseline model's accuracy by 10\% with the final macro accuracy being 84.8\%. 
	
	\section{Data}
	\label{sec:data}
	In this work, we use \textbf{TUT Acoustic Scenes 2017} dataset \cite{Mesaros2016_EUSIPCO} for development. The dataset is a collection of recordings from various acoustic scenes all from distinct locations.
	For each recording location 3-5 minute long audio recordings are captured and are split into 10 seconds which act as unit of sample for this task.
	All the audio clips are recorded with 44.1 kHz sampling rate and 24 bit resolution. Readers can find more information about the data and how it is collected from here\footnote{\url{http://www.cs.tut.fi/sgn/arg/dcase2017/challenge/task-acoustic-scene-classification}}. The development dataset comprises of 52 minutes of audio for each label amounting to 13 hours of entire dataset. On the contrary, the 2016 dataset comprised of over 9.75 hours of development dataset. Apart from the increase in dataset size, the datasets also differ in their classification units. The 2017 dataset has a classification unit of 10 seconds whereas the 2016 dataset has 30 seconds. These differences need to be observed before comparing the results from the 2017 edition with those of 2016 edition.
	
	\section{System Architecture}
	\label{sec:system_description}
	
	\begin{figure}[t]
		\centering
		\centerline{\includegraphics[width=\columnwidth]{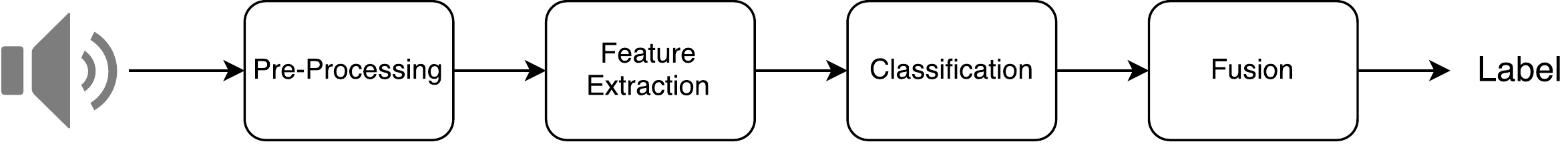}}
		\caption{Audio Processing Pipeline}
		\label{fig:results}
	\end{figure}
	
	For each audio clip, (from development and evaluation set), the processing pipeline consists of the following steps: 1) Pre-process the audio with various transformations to standardize the input, 2) Extract features from the pre-processed audio, 3) Train the DCNN models using the features from step 2, 4) To get the final classification for each file, we fuse the predictions from each segment. 
	
	In the following subsections, we explore each of these steps.
	
	\subsection{Audio Preprocessing}
	\label{ssec:preprocessing}
	Before being passed to the feature extractor, a few preprocessing steps
	are carried out on the original audio. In this task, the audio files have two sets of channels (one for left ear and one for right ear), so we convert the audio to mono by averaging the channels. We also normalize the amplitude to lie between -1 and 1.
	
	\subsection{Feature Extraction}
	\label{ssec:features}
	Spectrograms provide a visual way of representing the signal strength over time at various frequencies. Mel-Spectrograms maps the equally spaced spectrogram frequencies into bins according to human ear perception, hence using mel-spectrogram as input has shown lot of success in various tasks like speech recognition, speaker identification etc. In this task, we experiment with two variants of feature vectors: 
	
	\begin{enumerate}
		\item The original audio was downsampled to 16 kHz. We compute log mel-spectrogram with 0.025 seconds as window length and 0.010 seconds as hop length with 64 mel frequencies. The above processing results in log mel-spectrogram with 999 frames and 64 mel frequency bins. We then splits these into non overlapping windows of 111 frames which acts as input to our neural networks. Hence, we effectively split the raw audio into 9 segments.
		\item In this variant, the audio was kept in its original sampling frequency of 44.1 kHz. We compute log mel-spectrogram with 0.046 seconds as window length and 0.023 seconds as hop length with 64 mel frequencies. The above processing results in log mel-spectrogram with 431 frames and 64 mel frequency bins. We then splits these into non overlapping windows of 43 frames which acts as input to our neural networks. Hence, we effectively split the raw audio into 10 segments.
	\end{enumerate}
	
	\subsection{Classification}
	
	In this work, we apply a variety of DCNNs. They are described in detail in Section \ref{sec:cnn}. Once the feature vectors are extracted from the audio file, it is split into smaller segments as described in sub-section \ref{ssec:features}. Each model is trained using mini-batches of size 256. We used Adadelta optimizer \cite{zeiler2012adadelta} with an initial learning rate of 1. Dropout layers significantly improve performance. Hence, we added dropout rate of 0.5. All our models were trained for 200 epochs, and the models with best validation accuracy were chosen.
	
	\subsection{Fusion}
	Since we split the raw audio clip into multiple segments, to get the final class prediction, we average predictions across all segments of an audio clip.  This implies averaging  9 segments for the feature vector of type 1, and 10 segments for the feature vector of type 2 as mentioned in subsection \ref{ssec:features}.
	
	\subsection{Ensemble}
	
	Finally, we create an ensemble using the model variants mentioned in Section \ref{sec:cnn}. We trained 10 different models with 5 distinct architectures. We sort the models via the macro-accuracy computed using \cite{mesaros2016metrics}. Finally, we chose the best models which had the maximum variance, and whose accuracies were better than the baseline model. 
	
	We train our deep learning models with the Keras library \cite{chollet2015keras} using Tensorflow \cite{tensorflow2015-whitepaper} backend on cloud GPU's with 60 GB RAM hosted on Floydhub. \footnote{https://www.floydhub.com/}
	
	\section{Convolutional Neural Networks}
	\label{sec:cnn}
	
	Convolutional Neural Networks is yet another DNN which gained popularity in image classification tasks. DCNNs are an extension to DNN with a few changes to the architecture. The major difference is the usage of convolutions. A simple DCNN comprises of a set of layers stacked together. These are namely convolutional, pooling, optional fully connected layers at the end, and finally the output layer. 
	\begin{itemize}
	\item The convolutional layers consists of a kernel (also called as filter).  These kernels perform non-linear operations over a limited receptive field. The kernel is tiled across the entire input space, resulting in the creation of a feature map. The primary motivation for this operation is that the function which the kernel learns are independent of the position in which they are found. Hence, we reduce the number of parameters required for the neural network. The parameters for this layer are: context of the kernel (width x height), depth of the kernel and the strides. A typical convolutional layer will consist of numerous kernels. The depth depends on the number of kernels present in the previous layer. In the case of 1D convolution, the context of the kernel is defined only by its width. 
	\item The pooling layers are added to further reduce the dimensionality of the feature maps. The pooling layer essentially provide summaries over each context window, thus enhancing the network invariance to transitional shifts in the input patterns. Usually, pooling layers are placed after every convolution layer, however, in some DCNN variants, these are placed scantily. There are multiple types of pooling layers, max pooling, average pooling, global max pooling global average pooling, etc. The most salient feature of the pooling layer is that it doesn't add any additional parameters in the network and in turn helps reduce the dimensionality of the data significantly. Similar to the conv layer, pooling layer also have a context and a stride. In most typical scenarios, the strides are configured such that we pool over non-overlapping windows. 
	\end{itemize}
	This hierarchical structure consisting of alternating feature extraction layers and pooling layers allows CNNs to operate on multiple timescales. 
	In the upcoming sub-sections we describe the different DCNN architectures we've experimented with. Since the annotated dataset for the competition was small in size, we've used DCNN architectures which require less number of parameters. 
	
	\begin{table}[t]
	    \centering
	    \begin{tabular}{||c|c|c||}
	    \hline
 \textbf{LeNet} & \textbf{SqueezeNet} & \textbf{1D CNN} \\ \hline	    
	          8x3x3-BN-ReLu & 64x3x3-BN-ReLu &  64x5
	          -BN-ReLu\\\hline
	          MP: 3x2 & MP: 2x2 & MP: 3 \\ \hline
	          16x3x3-BN-ReLu &  FIRE(sq:16, ex:64) &  128x5-BN-ReLu\\\hline
	          
	          MP: 3x2 & FIRE(sq:16, ex:64) & MP: 3 \\ \hline
	          
	          32x3x3-BN-ReLu &  MP: 2x2 &  256x5-BN-ReLu\\\hline
	          Dropout: 0.5 & FIRE(sq:32, ex:128) & Dropout: 0.5  \\ \hline
	          FC: 512 & FIRE(sq:32, ex:128) & FC: 512 \\ \hline
	          Softmax: 15 & MP: 2x2 & Softmax: 15 \\ \hline
	          & FIRE(sq:48, ex:192) & \\ \hline
	          & FIRE(sq:64, ex:256) & \\ \hline
	          & Dropout: 0.5 & \\ \hline
	          & 15x1x1-BN-ReLu & \\ \hline
	          & GlobalAvgPool & \\ \hline

	    \end{tabular}
	    \caption{The various model architectures experimented in the paper and described in Section \ref{sec:cnn}. Some abbreviations used are:- BN: Batch Normalization, MP: MaxPooling Layer, FC: Fully Connected Layer, FIRE: fire module described in \ref{ssec:squeeze}, sq: squeeze, ex: expand.}
	    \label{tab:model_config}
	\end{table}
	
	\subsection{LeNet}
	We experimented with 3 variants of the LeNet architecture \cite{lecun1998gradient} by varying the filter sizes keeping everything else constant. We used the filter sizes of 3x3, 5x5, 7x7. A small building block of this architecture comprises of Conv-BN-Relu-Pool layer. We stacked 3 copies of the building blocks where the number of filters became twice of the previous block. Finally, we flatten the layer, add a fully connected layer of size 512 and use softmax for the final class label. The architecture is described in Table \ref{tab:model_config} for filter size: 3x3.

	\subsection{SqueezeNet} \label{ssec:squeeze}
	SqueezeNet \cite{SqueezeNet} is a DCNN which achieves AlexNet-level accuracy on ImageNet with 50x fewer parameters. SqueezeNet comprises of multiple stacked building blocks called Fire Modules. A Fire module is comprised of: a squeeze convolution layer (which has only 1x1 filters), feeding into an expand layer that has a mix of 1x1 and 3x3 convolution filters. The original SqueezeNet architecture comprises of 8 stacked Fire Modules where the number of filters keeps increasing as the depth increases. We modified the existing architecture by reducing the depth and stacking 6 Fire Modules instead. We experimented with various FIRE module configurations, the best configuration is described in Table \ref{tab:model_config}.
	
	\subsection{1D CNN}
	While applying CNN to audio dataset, it is important to understand that the two axes of the spectrogram have different meanings (time v/s frequency), which is essentially not the case for images. Hence, technically it makes sense to convolve only in a single dimension as opposed to both the dimensions in case of images. Hence, in this architecture we experiment with convolution in only single dimension (time). This architecture is inspired from \cite{NIPS2013_5004} \footnote{http://benanne.github.io/2014/08/05/spotify-cnns.html} with the major variation being, in the penultimate layer that variation uses various statistics, we used the entire layer and added a single hidden layer. This was primarily because we didn't want to overfit with large number of parameters.
	
	\section{Results and Analysis}
	\label{sec:results}

	From feature extraction side we experimented with sampling rate, window length and hop length when converting audio to log mel-spectrogram. If the window length is too big, frequency resolution with respect to time is reduced. If the window is too small, frequency patters across time are missed. This is a trade-off, one can have high a resolution in time or a high resolution in frequency but not both. So, we experimented with various window and hop lengths. The human ear can listen to frequencies between 0Hz and 5kHz clearly without any effort, so, it is a common practice to downsample the audio. Hence, we  experimented with the original as well as the downsampled audio files. 
	
	On the model architecture side, we experimented with LeNet, SqueezeNet and 1D CNNs with different kernel sizes. In LeNet we used filters of size 3x3, 5x5 and 7x7 to understand how filter size affects the classification accuracy.
	
	The model \textbf{CNN-V1} is trained on log mel-spectrograms of audio files without down-sampling with 3x3 conv filters whereas all the other models are trained on log mel-spectrograms of down-sampled (16kHz) audio files. While training we observed that the model overfits very quickly when using actual files because it contains lot of information which may not be relevant to task at hand results in poor accuracy. The models trained on down-sampled audio files are robust to overfitting and hence perform better. 
	
	The models \textbf{CNN-V2-1}, \textbf{CNN-V2-2} and \textbf{CNN-V2-3} are based on LeNet architecture, where the only difference is the size of the conv filters used. Conv filters of size 3x3, 5x5 and 7x7 are used in them respectively. The conv filter sizes affect the model performance. The filters which are either too big or too small cannot understand the time-frequency patterns. Hence, we experimented with various filter sizes and we can see that the classification accuracy deteriorates with the  increase in the filter size. Due to recent success of \textbf{SqueezNet} with very less number of parameters we tried this architecture out of curiosity.
	
	We finally tried \textbf{1D CNN} because a spectrogram is not like an image, it represents frequency in one axis and time in one axis. Even though we can use it like an image sometimes model may benefit from treating them separately.
	
	After training all the models we create an ensemble by choose best three models and combined their individual predictions using geometric mean. Since the models are learning different because of their architectures (3x3 vs 5x5 etc) or input features (16kHz-44.1kHz log mel-spectrogram) we see a improvement in accuracy of about 3 percent compared to best individual model.
	
	We plotted confusion matrix in figure \ref{fig:conf_matrix} to analyze the conflicting acoustic scenes. The most confused pairs are: \textbf{Home} is confused with \textbf{Library}, and \textbf{Park} is confused with \textbf{Residential area}. We can clearly see why the model is struggling to identify these pairs because these scenes are closely related to one another acoustically.
	
	This can be reduced by increasing the audio sample length from 10 seconds so that we have some more context to identify these pairs correctly. Among the most accurately predicted acoustic scenes there are \textbf{office}, \textbf{car}, \textbf{city\_center}. These scenes have unique fingerprint to them and could be accurately classified. On the other hand, \textbf{park}, \textbf{cafe restaurant}, \textbf{residential\_area} are least accurately predicted probably due to acoustic closeness of these scenes. 
	
	\begin{table*}[!htbp]
	\centering
		{
			\begin{tabular}{||c|c|c|c|c|c|c|c|c||}
				\hline
				\textbf{Acoustic Scene} & \textbf{Base} & \textbf{CNN-V1}  & \textbf{CNN-V2-1} & \textbf{CNN-V2-2} & \textbf{CNN-V2-3} & \textbf{SqueezeNet} & \textbf{1D CNN} & \textbf{Ensemble}\\
						
				\hline \hline
						
				Beach             & 75.3 - 40.7 & 87.2 - 13.0 & 83.3 - 8.3 & 83.7 - 10.2 & 85.3 & 84.6 & 75.0 & 87.5 - 13.9 \\ \hline
				Bus               & 75.3 - 38.9 & 93.9 - 35.2 & 84.9 - 39.8 & 92.9 - 49.1 & 86.9 & 85.9 & 84.3 & 90.1 - 42.6 \\  \hline
				Cafe / Restaurant & 57.7 - 43.5 & 66.0 - 51.9 & 72.1 - 57.4 & 64.4 - 45.4 & 64.1 & 53.2 & 55.8 & 72.8 - 57.4 \\   \hline
				Car               & 97.1 - 64.8 & 94.6 - 88.0 & 97.8 - 96.3 & 94.2 - 77.8 & 97.4 & 93.6 & 97.8 & 98.4 - 85.2 \\   \hline
				City center       & 90.7 - 79.6 & 82.1 - 85.2 & 90.1 - 75.9 & 91.7 - 89.8 & 93.6 & 84.9 & 83.7 & 93.6 - 85.2 \\   \hline
				Forest path       & 79.5 - 85.2 & 85.9 - 86.1 & 87.5 - 88.0 & 93.6 - 85.2 & 91.3 & 82.4 & 76.0 & 91.7 - 87.0 \\   \hline
				Grocery store     & 58.7 - 49.1 & 79.2 - 52.8 & 90.1 - 58.3 & 90.1 - 54.6 & 85.6 & 87.2 & 89.4 & 92.9 - 57.4 \\   \hline
				Home              & 68.6 - 76.9 & 70.1 - 68.5 & 74.5 - 79.6 & 74.2 - 81.5 & 69.5 & 77.7 & 72.0 & 74.8 - 83.3 \\   \hline
				Library           & 57.1 - 30.6 & 81.7 - 25.0 & 81.7 - 34.3 & 85.9 - 38.9 & 78.8 & 73.7 & 77.2 & 87.2 - 35.2 \\   \hline
				Metro station     & 91.7 - 93.5 & 93.6 - 28.7 & 69.9 - 23.1 & 71.8 - 97.2 & 65.1 & 67.9 & 69.9 & 86.5 - 63.9 \\   \hline
				Office            & 99.7 - 73.1 & 92.6 - 72.2 & 99.7 - 86.1 & 99.0 - 94.4 & 99.0 & 97.8 & 98.1 & 98.4 - 88.9 \\   \hline
				Park              & 70.2 - 32.4 & 60.6 - 35.2 & 70.8 - 40.7 & 62.8 - 25.0 & 59.9 & 56.1 & 60.9 & 66.3 - 31.5 \\   \hline
				Residential area  & 64.1 - 77.8 & 70.2 - 82.4 & 70.8 - 78.7 & 75.6 - 80.6 & 67.9 & 72.4 & 69.6 & 73.4 - 81.5 \\   \hline
				Train             & 58.0 - 72.2 & 61.9 - 71.3 & 74.4 - 74.1 & 66.3 - 75.0 & 67.3 & 42.0 & 55.1 & 75.3 - 72.2 \\   \hline
				Tram              & 81.7 - 57.4 & 78.5 - 60.2 & 80.4 - 57.4 & 77.9 - 56.5 & 81.1 & 77.2 & 82.1 & 83.3 - 60.2 \\   \hline
				Average Accuracy  & 74.8 - 61.0 & 79.9 - 57.0 & 81.9 - 59.9 & 81.6 - 64.1 & 79.5 & 75.8 & 76.4 & 84.8 - 63.0 \\ \hline \hline
								    
			\end{tabular}}
		\caption{\label{results} Class-wise accuracy of acoustic scenes for various cnn models we experimented with. Refer Section \ref{sec:results} for details on model notation. Results with a pair of number represent \textbf{development accuracy - evaluation accuracy} whereas the ones with a single number represent just development accuracy.}
	\end{table*}

	\begin{figure}[!htbp]
        \centering 
        \includegraphics[width=\columnwidth]{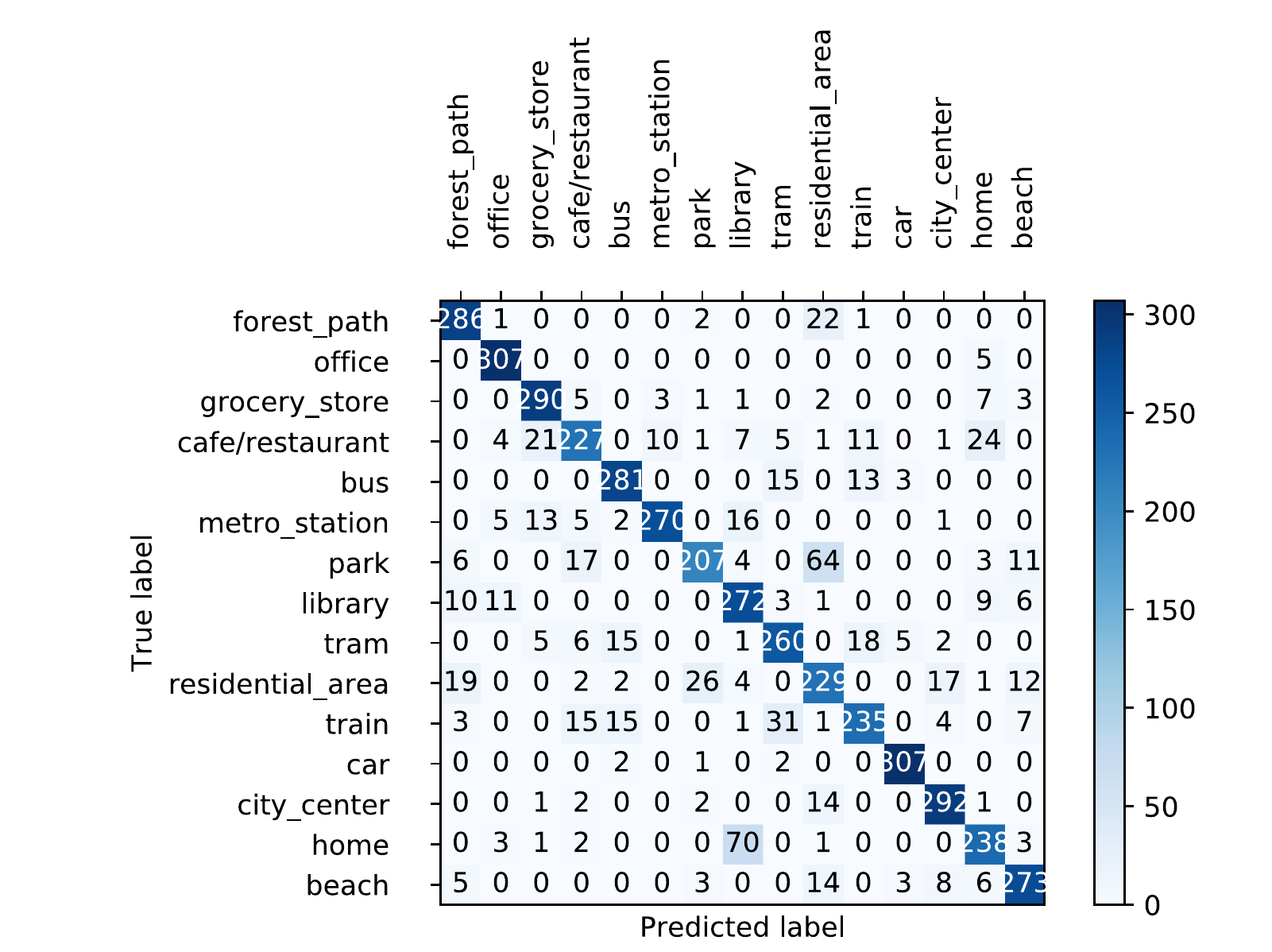}
        \caption{Confusion Matrix}
        \label{fig:conf_matrix}
    \end{figure}

	\section{Future Work \& Conclusion}
	Currently, the training suffers because of over fitting due to lack of enough data. In future we'll try to avoid this using data augmentation methods as these methods have shown to boost the performance and prevent over-fitting in \cite{han2016acoustic} \cite{salamon2017deep}. In this work, we experimented with the window length and the hop length of mel-spectrogram to study its affect on acoustic scene classification. We also tried different DCNN architectures for the same and finally created an ensemble of these deep CNN models to improve upon the baseline model by 10\%.
	
	\bibliographystyle{IEEEtran}
	\bibliography{refs}

\begin{thebibliography}{10}
\providecommand{\url}[1]{#1}
\def\UrlFont{\rmfamily}
\providecommand{\newblock}{\relax}
\providecommand{\bibinfo}[2]{#2}
\providecommand\BIBentrySTDinterwordspacing{\spaceskip=0pt\relax}
\providecommand\BIBentryALTinterwordstretchfactor{4}
\providecommand\BIBentryALTinterwordspacing{\spaceskip=\fontdimen2\font plus
\BIBentryALTinterwordstretchfactor\fontdimen3\font minus
  \fontdimen4\font\relax}
\providecommand\BIBforeignlanguage[2]{{%
\expandafter\ifx\csname l@#1\endcsname\relax
\typeout{** WARNING: IEEEtran.bst: No hyphenation pattern has been}%
\typeout{** loaded for the language `#1'. Using the pattern for}%
\typeout{** the default language instead.}%
\else
\language=\csname l@#1\endcsname
\fi
#2}}

\bibitem{barchiesi2015acoustic}
D.~Barchiesi, D.~Giannoulis, D.~Stowell, and M.~D. Plumbley, ``Acoustic scene
  classification: Classifying environments from the sounds they produce,''
  \emph{IEEE Signal Processing Magazine}, vol.~32, no.~3, pp. 16--34, 2015.

\bibitem{wang2006computational}
D.~Wang and G.~J. Brown, \emph{Computational auditory scene analysis:
  Principles, algorithms, and applications}.\hskip 1em plus 0.5em minus
  0.4em\relax Wiley-IEEE press, 2006.

\bibitem{cai2006flexible}
R.~Cai, L.~Lu, A.~Hanjalic, H.-J. Zhang, and L.-H. Cai, ``A flexible framework
  for key audio effects detection and auditory context inference,'' \emph{IEEE
  Transactions on audio, speech, and language processing}, vol.~14, no.~3, pp.
  1026--1039, 2006.

\bibitem{shah2012lifelogging}
M.~Shah, B.~Mears, C.~Chakrabarti, and A.~Spanias, ``Lifelogging: Archival and
  retrieval of continuously recorded audio using wearable devices,'' in
  \emph{Emerging Signal Processing Applications (ESPA), 2012 IEEE International
  Conference on}.\hskip 1em plus 0.5em minus 0.4em\relax IEEE, 2012, pp.
  99--102.

\bibitem{chu2006scene}
S.~Chu, S.~Narayanan, C.-C.~J. Kuo, and M.~J. Mataric, ``Where am i? scene
  recognition for mobile robots using audio features,'' in \emph{Multimedia and
  Expo, 2006 IEEE International Conference on}.\hskip 1em plus 0.5em minus
  0.4em\relax IEEE, 2006, pp. 885--888.

\bibitem{deng2009imagenet}
J.~Deng, W.~Dong, R.~Socher, L.-J. Li, K.~Li, and L.~Fei-Fei, ``Imagenet: A
  large-scale hierarchical image database,'' in \emph{Computer Vision and
  Pattern Recognition, 2009. CVPR 2009. IEEE Conference on}.\hskip 1em plus
  0.5em minus 0.4em\relax IEEE, 2009, pp. 248--255.

\bibitem{giannoulis2013detection}
D.~Giannoulis, E.~Benetos, D.~Stowell, M.~Rossignol, M.~Lagrange, and M.~D.
  Plumbley, ``Detection and classification of acoustic scenes and events: An
  ieee aasp challenge,'' in \emph{Applications of Signal Processing to Audio
  and Acoustics (WASPAA), 2013 IEEE Workshop on}.\hskip 1em plus 0.5em minus
  0.4em\relax IEEE, 2013, pp. 1--4.

\bibitem{heittola2016dcase2016}
T.~Heittola, A.~Mesaros, and T.~Virtanen, ``Dcase2016 baseline system,'' 2016.

\bibitem{DCASE2017challenge}
A.~Mesaros, T.~Heittola, A.~Diment, B.~Elizalde, A.~Shah, E.~Vincent, B.~Raj,
  and T.~Virtanen, ``{DCASE} 2017 challenge setup: Tasks, datasets and baseline
  system,'' in \emph{Proceedings of the Detection and Classification of
  Acoustic Scenes and Events 2017 Workshop (DCASE2017)}, November 2017,
  submitted.

\bibitem{chum2013ieee}
M.~Chum, A.~Habshush, A.~Rahman, and C.~Sang, ``Ieee aasp scene classification
  challenge using hidden markov models and frame based classification,''
  \emph{IEEE AASP Challenge on Detection and Classification of Acoustic Scenes
  and Events}, 2013.

\bibitem{olivetti2013wonders}
E.~Olivetti, ``The wonders of the normalized compression dissimilarity
  representation,'' \emph{IEEE AASP Challenge on Detection and Classification
  of Acoustic Scenes and Events}, 2013.

\bibitem{geiger2013large}
J.~T. Geiger, B.~Schuller, and G.~Rigoll, ``Large-scale audio feature
  extraction and svm for acoustic scene classification,'' in \emph{Applications
  of Signal Processing to Audio and Acoustics (WASPAA), 2013 IEEE Workshop
  on}.\hskip 1em plus 0.5em minus 0.4em\relax IEEE, 2013, pp. 1--4.

\bibitem{eghbal2016cp}
H.~Eghbal-Zadeh, B.~Lehner, M.~Dorfer, and G.~Widmer, ``Cp-jku submissions for
  dcase-2016: A hybrid approach using binaural i-vectors and deep convolutional
  neural networks,'' \emph{IEEE AASP Challenge on Detection and Classification
  of Acoustic Scenes and Events (DCASE)}, 2016.

\bibitem{valenti2016dcase}
M.~Valenti, A.~Diment, G.~Parascandolo, S.~Squartini, and T.~Virtanen, ``Dcase
  2016 acoustic scene classification using convolutional neural networks,'' in
  \emph{Proc. Workshop Detection Classif. Acoust. Scenes Events}, 2016, pp.
  95--99.

\bibitem{lidy2016cqt}
T.~Lidy and A.~Schindler, ``Cqt-based convolutional neural networks for audio
  scene classification and domestic audio tagging,'' \emph{IEEE AASP Challenge
  on Detection and Classification of Acoustic Scenes and Events (DCASE 2016),
  Budapest, Hungary, Tech. Rep}, 2016.

\bibitem{bae2016acoustic}
S.~H. Bae, I.~Choi, and N.~S. Kim, ``Acoustic scene classification using
  parallel combination of lstm and cnn,'' in \emph{Proceedings of the Detection
  and Classification of Acoustic Scenes and Events 2016 Workshop (DCASE2016)},
  2016.

\bibitem{Mesaros2016_EUSIPCO}
A.~Mesaros, T.~Heittola, and T.~Virtanen, ``{TUT} database for acoustic scene
  classification and sound event detection,'' in \emph{24th European Signal
  Processing Conference 2016 (EUSIPCO 2016)}, Budapest, Hungary, 2016.

\bibitem{zeiler2012adadelta}
M.~D. Zeiler, ``Adadelta: an adaptive learning rate method,'' \emph{arXiv
  preprint arXiv:1212.5701}, 2012.

\bibitem{mesaros2016metrics}
A.~Mesaros, T.~Heittola, and T.~Virtanen, ``Metrics for polyphonic sound event
  detection,'' \emph{Applied Sciences}, vol.~6, no.~6, p. 162, 2016.

\bibitem{chollet2015keras}
F.~Chollet \emph{et~al.}, ``Keras,'' \url{https://github.com/fchollet/keras},
  2015.

\bibitem{tensorflow2015-whitepaper}
\BIBentryALTinterwordspacing
M.~Abadi, A.~Agarwal, P.~Barham, E.~Brevdo, Z.~Chen, C.~Citro, G.~S. Corrado,
  A.~Davis, J.~Dean, M.~Devin, S.~Ghemawat, I.~Goodfellow, A.~Harp, G.~Irving,
  M.~Isard, Y.~Jia, R.~Jozefowicz, L.~Kaiser, M.~Kudlur, J.~Levenberg,
  D.~Man\'{e}, R.~Monga, S.~Moore, D.~Murray, C.~Olah, M.~Schuster, J.~Shlens,
  B.~Steiner, I.~Sutskever, K.~Talwar, P.~Tucker, V.~Vanhoucke, V.~Vasudevan,
  F.~Vi\'{e}gas, O.~Vinyals, P.~Warden, M.~Wattenberg, M.~Wicke, Y.~Yu, and
  X.~Zheng, ``{TensorFlow}: Large-scale machine learning on heterogeneous
  systems,'' 2015, software available from tensorflow.org. [Online]. Available:
  \url{http://tensorflow.org/}
\BIBentrySTDinterwordspacing

\bibitem{lecun1998gradient}
Y.~LeCun, L.~Bottou, Y.~Bengio, and P.~Haffner, ``Gradient-based learning
  applied to document recognition,'' \emph{Proceedings of the IEEE}, vol.~86,
  no.~11, pp. 2278--2324, 1998.

\bibitem{SqueezeNet}
F.~N. Iandola, S.~Han, M.~W. Moskewicz, K.~Ashraf, W.~J. Dally, and K.~Keutzer,
  ``Squeezenet: Alexnet-level accuracy with 50x fewer parameters and $<$0.5mb
  model size,'' \emph{arXiv:1602.07360}, 2016.

\bibitem{NIPS2013_5004}
\BIBentryALTinterwordspacing
A.~van~den Oord, S.~Dieleman, and B.~Schrauwen, ``Deep content-based music
  recommendation,'' in \emph{Advances in Neural Information Processing Systems
  26}, C.~J.~C. Burges, L.~Bottou, M.~Welling, Z.~Ghahramani, and K.~Q.
  Weinberger, Eds.\hskip 1em plus 0.5em minus 0.4em\relax Curran Associates,
  Inc., 2013, pp. 2643--2651. [Online]. Available:
  \url{http://papers.nips.cc/paper/5004-deep-content-based-music-recommendation.pdf}
\BIBentrySTDinterwordspacing

\bibitem{han2016acoustic}
Y.~Han and K.~Lee, ``Acoustic scene classification using convolutional neural
  network and multiple-width frequency-delta data augmentation,'' \emph{arXiv
  preprint arXiv:1607.02383}, 2016.

\bibitem{salamon2017deep}
J.~Salamon and J.~P. Bello, ``Deep convolutional neural networks and data
  augmentation for environmental sound classification,'' \emph{IEEE Signal
  Processing Letters}, vol.~24, no.~3, pp. 279--283, 2017.

\end{thebibliography}
	%
	%
	%
	%
	%
	%
	%
	%
	%

\end{sloppy}
\end{document}